\documentclass{aa}

\usepackage{graphics}

\begin{document}

   \thesaurus{11(05.01.1; 08.02.3; 08.22.1)}

   \title{Polaris: astrometric orbit, position, and proper motion}

   \subtitle{}

   \author{R. Wielen
   \and H. Jahrei{\ss}
   \and C. Dettbarn
   \and H. Lenhardt
   \and H. Schwan}

   \offprints{R. Wielen (wielen@ari.uni-heidelberg.de)}

   \institute{Astronomisches Rechen-Institut, Moenchhofstrasse 12-14,
   D-69120 Heidelberg, Germany}

   \date{Received 14 February 2000 / Accepted 22 April 2000}

\authorrunning{R. Wielen et al.}

\titlerunning
{Polaris: astrometric orbit, position, and proper motion}

\maketitle

\begin{abstract}
We derive the astrometric orbit of the photo-center of the close pair
$\alpha$ UMi AP (= $\alpha$ UMi Aa)
of the Polaris multiple stellar system. The orbit is based on the
spectroscopic orbit of the Cepheid $\alpha$ UMi A (orbital period of AP: 29.59
years), and on the difference $\Delta\mu$ between the quasi-instantaneously
measured HIPPARCOS proper motion of Polaris and the long-term-averaged proper
motion given by the FK5. There remains an ambiguity in the inclination $i$ of
the orbit, since $\Delta\mu$ cannot distinguish between a prograde orbit $(i =
50\,.\!\!^\circ1)$ and a retrograde one $(i = 130\,.\!\!^\circ2)$. Available
photographic observations of Polaris favour strongly the retrograde orbit. For
the semi-major axis of the photo-center of AP we find about
29 milliarcsec (mas). For the
component P, we estimate a mass of 1.5 ${\cal M}_\odot$ and a magnitude
difference with respect to the Cepheid of 6.5 mag. The present separation
between A and P should be about 160 mas.

We obtain the proper motion of the center-of-mass of $\alpha$ UMi AP with a
mean error of about 0.45 mas/year. Using the derived astrometric orbit, we find
the position of the center-of-mass at the epoch 1991.31 with an
accuracy of about 3.0 mas. Our ephemerides for the orbital correction, required
for going from the position of the center-of-mass to the instantaneous position
of the photo-center of AP at an arbitrary epoch, have
a typical uncertainty of 5
mas. For epochs which differ from the HIPPARCOS epoch by more than a few years,
a prediction for the actual position of Polaris based on our results should be
significantly more accurate than using the HIPPARCOS data in a linear
prediction, since the HIPPARCOS proper motion contains the instantaneous
orbital motion of about 4.9 mas/year = 3.1 km/s. Finally we derive the galactic
space motion of Polaris.

\keywords{astrometry -- binaries: general -- Cepheids}
\end{abstract}

\section{Introduction}

\begin{figure}[t]
\resizebox{\hsize}{!}{\includegraphics{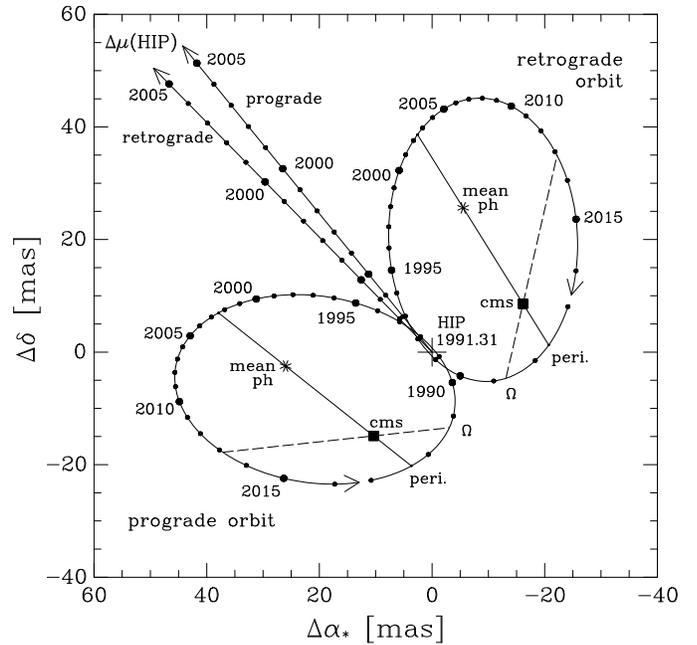}}
\caption[]
{Astrometric orbit (prograde or retrograde) of the photo-center of $\alpha$
UMi AP. The retrograde orbit is our preferred solution. For detailed
explanations see Sect. 3.2.4\,.}
\end{figure}

Polaris ($\alpha$ Ursae Minoris, HR 424, HD 8890, ADS 1477, FK 907, HIP 11767)
is a very interesting and important
object, both from the astrophysical point of view and
from the astrometric one. For astrophysics, the most remarkable
feature of the multiple stellar system Polaris is the fact that its main
component, namely $\alpha$ UMi A, is a Cepheid variable with a very unusual
behaviour. In astrometry, Polaris is one of the most frequently and accurately
observed objects, mainly because it is located so close to the North celestial
pole and can be used for calibration purposes.

Up to now, the binary nature of Polaris was essentially neglected in
ground-based fundamental astrometry, e.g in the FK5 (Fricke et al. 1988). This
was justified
by the limited accuracy reached by the meridian-circle observations. Now, the
high-precision astrometric measurements carried out with the HIPPARCOS
satellite (ESA 1997) require strongly to take into account the binary nature of
Polaris in order to obtain an adequate
astrometric description of $\alpha$ UMi. Similar procedures are required for
many other binaries among the fundamental stars in order to be included
properly into the Sixth Catalogue of Fundamental Stars (FK6; Part I: Wielen et
al. 1999c; see also Wielen et al. 1998).

The main purpose of the present paper is to obtain a reliable astrometric orbit
for Polaris, and to use this astrometric orbit for obtaining high-precision
values for the position and proper motion of Polaris. This is done by combining
the known spectroscopic orbit of Polaris with ground-based astrometric data
given in the FK5 and with the HIPPARCOS results. Before doing so in the
Sects. 3 and 4, we present in Sect. 2 an overview of the Polaris
system.

\section{Overview of the Polaris System}

Polaris is a multiple stellar system, which consists of a close pair, $\alpha$
UMi A and $\alpha$ UMi P (= $\alpha$ UMi a),
and a distant companion, $\alpha$ UMi B, and two
distant components, $\alpha$ UMi C and $\alpha$ UMi D.
We use here the designation 'P' for a close companion of A,
which was used in the IDS and was adopted by the CCDM and by the
HIPPARCOS Input Catalogue,
rather than the traditional version 'a', which is
used e.g. by the WDS and by CHARA).

\subsection{The Cepheid $\alpha$ UMi A}

The main component of Polaris is a low-amplitude Cepheid with a pulsational
period of about 3.97 days. This period is increasing with time (e.g. Kamper \&
Fernie 1998). According to Feast \& Catchpole (1997), $\alpha$ UMi A is a
first-overtone pulsator (rather than a fundamental one),
since $\alpha$ UMi A is too luminous for a fundamental pulsator,
if they apply their period-luminosity relation (for fundamental pulsators)
to Polaris. The fundamental period of $\alpha$ UMi A would follow as
$P_0 = 5.64$ days, if the observed period is the first-overtone period $P_1$
(using the relation between
$P_1$ and $P_0$ derived by Alcock et al. (1995) for Galactic Cepheids).
An extraordinary property
of $\alpha$ UMi A among
the Cepheids is that the amplitude of its pulsation has
been dramatically declined during the past 100 years, as seen both in the light
curve and in the radial-velocity curve (Arellano Ferro 1983, Kamper \& Fernie
1998, and other references given therein). The full amplitude was
about 0\,.$\!\!^m$12 in $m_{\rm V}$ and about 6 km/s in radial velocity
before 1900, and seems now to be rather constant
at a level of only
0\,.$\!\!^m$03 in $m_{\rm V}$ and at 1.6 km/s in radial velocity. An
earlier prediction (Fernie et al. 1993) that the pulsation should cease totally
in the 1990s was invalid. A discussion of the HIPPARCOS parallax and of the
absolute magnitude of $\alpha$ UMi A is given in the next Sect. 2.2\,.

\subsection{The spectroscopic-astrometric binary $\alpha$ UMi AP}

The Cepheid $\alpha$ UMi A is a member of the close binary system $\alpha$ UMi
AP. This duplicity was first found from the corresponding variations in the
radial velocity of $\alpha$ UMi A. However, the interpretation of the radial
velocities of $\alpha$ UMi A in terms of a spectroscopic binary is obviously
complicated by the fact that $\alpha$ UMi A itself is pulsating and that this
pulsation varies with time. We use in this paper the spectroscopic orbit
derived by Kamper (1996), which is based on radial velocity observations from
1896 to 1995. Kamper (1996) took into account changes in the amplitude of the
pulsation and in the period of pulsation, but used otherwise a fixed sinusoid
for fitting the pulsation curve. In an earlier paper, Roemer (1965) considered
even `annual' changes in the form of the pulsation curve. In Table 4, we list
the elements of the spectroscopic orbit of A in the pair AP given by Kamper
(1996, his Table III, DDO + Lick Data). The orbital period of $\alpha$ UMi AP
is 29.59\,$\pm$\,0.02 years, and the semi-amplitude is
$K_{\rm A} = 3.72$ km/s.
The value of $a_{\rm A} \sin i = 2.934$ AU
corresponds to about
22 milliarcsec (mas), using the
HIPPARCOS parallax.

Attempts to observe the secondary component $\alpha$ UMi P directly or in the
integrated spectrum of $\alpha$ UMi AP have failed up to now. Burnham (1894)
examined Polaris in 1889 with the 36-inch Lick refractor and found no close
companion to $\alpha$ UMi A (nor to $\alpha$ UMi B).  Wilson (1937)
claimed to have observed a close companion by means of an interferometer
attached to the 18-inch refractor of the Flower Observatory. Jeffers
(according to Roemer (1965) and to the WDS Catalogue) was unable to confirm
such a companion
with an interferometer at the 36-inch refractor of the Lick Observatory.
HIPPARCOS (ESA 1997) has not given any indication for the duplicity of
Polaris. Speckle observations were also
unsuccessful (McAlister 1978). All these failures to detect
$\alpha$ UMi P directly are not astonishing in view of the probable magnitude
difference of A and P of more than 6$^m$ and a separation of A and P
of less than 0\,.$\!\!$"2 (see Sect. 3.2.5).
Roemer and Herbig (Roemer 1965) and Evans
(1988) searched without success for light from $\alpha$ UMi P in the combined
spectrum of $\alpha$ UMi AP. From IUE spectra, Evans (1988) concluded that a
main-sequence companion must be later than A8V. This is in agreement with our
results for $\alpha$ UMi P, given in Table 5. A white-dwarf companion is
ruled out by the upper limit on its effective temperature derived from IUE
spectra and by considerations on its cooling age, which would be much higher
than the age of the Cepheid $\alpha$ UMi A (Landsman et al. 1996).

After Polaris had become known as a long-period
spectroscopic binary (Moore 1929), various
attempts have been made to obtain an astrometric orbit for the pair $\alpha$
UMi AP. Meridian-circle observations were discussed by Gerasimovic (1936) and
van Herk (1939).  While van Herk did not find a regular variation with a
period of 30 years, Gerasimovic claimed to have found such a modulation.
However, the astrometric orbit of the visual photo-center of
$\alpha$ UMi AP determined by Gerasimovich
(1936) is most probably spurious, since he found for the semi-major axis of the
orbit $a_{{\rm ph (AP)}} \sim 110$ mas,
which is much too high in view of our
present knowledge
($a_{{\rm ph (AP)}} = 29$ mas). More recent meridian-circle
observations gave no indications of any significant perturbation. This is not
astonishing in view of the small orbital displacements of the
photo-center of AP of always less than 0\,.$\!\!$"04. Long-focus photographic
observations have
been carried out at the Allegheny Observatory (during 1922--1964), the
Greenwich Observatory, and the
Sproul Observatory (during 1926--1956),
mainly with the aim to determine the parallax of Polaris.
The discussion of
this material by Wyller (1957, Sproul data) and by Roemer (1965, Allegheny
data) did not produce any significant results. The Allegheny plates were later
remeasured and rediscussed by Kamper (1996), using his new spectroscopic
orbital elements. Kamper also rediscussed the Sproul plates. While the Sproul
data gave no relevant results for $\alpha$ UMi AP, the Allegheny data gave just
barely significant results, such as
$a_{{\rm ph (AP)}} = 19.5 \pm 6.5$ mas.
For our purpose (see Sect. 3.2.3), the most important implication derived by
Kamper (1996) from
the Allegheny data is that the astrometric orbit of AP is most
probably retrograde, not prograde.

In Sect. 3 we shall present a more reliable astrometric
orbit of $\alpha$ UMi AP by combining ground-based FK5 data with HIPPARCOS
results, using Kamper's (1996) spectroscopic orbit as a basis.

The HIPPARCOS astrometric satellite has obtained for $\alpha$ UMi AP a
trigonometric parallax of
$p_{\rm H} = 7.56 \pm 0.48$ mas, which corresponds
to a distance from the Sun of
$r_{\rm H} = 132 \pm 8$ pc. In the data
reduction
for HIPPARCOS, it was implicitely assumed that the photo-center of the pair AP
moves linearly in space and time, i.e. a `standard solution' was adopted. This
is a fairly valid assumption, since the deviations from a linear fit over the
period of observations by HIPPARCOS, about 3 years, are less than 1 mas (see
Sect. 4.2). Hence the HIPPARCOS parallax obtained is most probably not
significantly affected by the curvature of the orbit of AP. Nevertheless, it
may be reassuring to repeat the data reduction of HIPPARCOS for $\alpha$ UMi,
adopting the astrometric orbit derived here for implementing the curvature of
the orbit of the photo-center of $\alpha$ UMi AP.

The mean apparent visual magnitude of the combined components A and P is
$m_{{\rm V, AP}} = 1.982$ (Feast \& Catchpole 1997).
This agrees fairly well with the HIPPARCOS result (ESA 1997)
$m_{{\rm V, AP}} = 1.97 $.
In accordance with most authors we assume that the reddening $E_{\rm{B-V}}$
and the extinction $A_{\rm V}$ of the Polaris system are essentially zero
( e.g., Turner 1977, Gauthier \& Fernie 1978),
within a margin of $\pm 0.02$ in $E_{\rm{B-V}}$
and $\pm 0.06$ in $A_{\rm V}$. Using
the HIPPARCOS parallax, we find for the mean absolute magnitude of AP
$M_{{\rm V, AP}} = - 3.63 \pm 0.14$.
If we use our results of Table 5
for component P, i.e. $M_{{\rm V, P}} \sim + 2.9$,
and subtract the light of P from $M_{{\rm V, AP}}$,
then the absolute magnitude of the Cepheid component A is
$M_{{\rm V, A}} = -3.62 \pm 0.14$.
Unfortunately, the pecularities in the pulsation of $\alpha$ UMi
A are certainly not very favourable for using this nearest Cepheid as the main
calibrator of the zero-point of the period-luminosity relation of classical
Cepheids.

\subsection{The visual binary $\alpha$ UMi (AP)\,--\,B}

Already in 1779, W. Herschel (1782) discovered the visual-binary nature of
Polaris. The
present separation between AP and B is about 18\,.$\!\!$"2. This separation
corresponds to 2400 AU or 0.012 pc, if B has the same parallax as AP. Kamper
(1996) has determined the tangential and radial velocity of B relative to AP.
Both velocities of B agree with those of AP within about 1 km/s. Hence Kamper
(1996) concludes that B is most probably a physical companion of AP, and not an
optical component. The physical association between AP
and B is also supported by the fair agreement between the HIPPARCOS parallax of
AP ($r_{\rm H} = 132 \pm 8$ pc)
and the spectroscopic parallax of B (114
pc, as mentioned below).

The spectral type of B is F3V. The magnitude difference between B and the
combined light of AP is
$\Delta m_{\rm V} = 6.61 \pm 0.04$ (Kamper 1996).
Using $m_{{\rm V, AP}} = 1.98$, this implies for B an apparent magnitude
of $m_{{\rm V, B}} = 8.59 \pm 0.04$.
Adopting the HIPPARCOS parallax (and no extinction),
we obtain for B an absolute magnitude of
$M_{{\rm V, B}} = +2.98 \pm 0.15$.
The standard value of $M_{\rm V}$
for an F3V star on the zero-age main sequence
is $+ 3.3 $. If we use this standard value for $M_{\rm V}$, we obtain for B
a spectroscopic distance of $r = 114$ pc. Similar values of the
spectroscopic distance were derived (or implied) by Fernie (1966),
Turner (1977), and Gauthier \& Fernie (1978).
These authors were interested in the
absolute magnitude (and hence in the distance) of B
in order to calibrate the absolute magnitude
of the Cepheid A. Now the use of
the HIPPARCOS trigonometric parallax is, of course, better suited for this
purpose.

The typical mass of an F3V star is
${\cal M}_{\rm B} = 1.5 \, {\cal M}_{\odot}$. If
we use for the masses of A and P the
values adopted in Table 5 ($6.0 + 1.54 \, {\cal M}_\odot$), we
obtain for the triple system a total mass of ${\cal M}_{\rm{tot}} = 9.0 \,
{\cal M}_{\odot}$). We derive from
$\rho_{\rm{B-AP}} = 18\,.\!\!$"$2$ and the
statistical
relation $a = 1.13 \, \rho$ an estimate for the semi-major axis of the orbit
of B relative to AP of
$a_{\rm{B - AP}} \sim 21$" or 2700 AU. From Kepler's Third
Law, we get then an
estimate of the orbital period of B,
namely $P_{\rm B} \sim 50\,000$ years.

From the data given above, we can estimate the acceleration
$g_{{\rm AP}}$ of the
center-of-mass of the pair $\alpha$ UMi AP due to the gravitational attraction
of $\alpha$ UMi B. If we project this estimate of
$g_{{\rm AP}}$ on one arbitrarly
chosen direction, we get for AP a typical `one-dimensional' acceleration of
about 0.003 (km/s)/century or 0.4 mas/century$^2$. Therefore, we should expect
neither in the radial velocity nor in the tangential motion of AP
a significant deviation from linear motion due to the gravitational force of B
during the relevant periods of the observations used.
For all present
purposes, it is fully adequate to assume that the center-of-mass of the pair
$\alpha$ UMi AP moves linearly in space and time.
The same is true for the motion of B.

A modulation of the relative position of B with respect
to the photo-center of AP with a period of about 30 years
is not seen in the available observations of B.
This is in accordance with our determination of the motion
of the photo-center of AP with respect to the cms of AP, given in Table 7.
The expected amplitude of the modulation is
less than 0\,.$\!\!$"04 and is obviously not large enough with respect
to the typical measuring errors in the relative position of B.

The contribution of the orbital motion of the center-of-mass (cms) of AP, due
to B, to the total space
velocity of AP is of the order of a few tenth of a km/s. The expected value of
the velocity of B relative to the cms of AP is of the order of 1 km/s.


\begin{table*}

\caption[]{Mean proper motion of the photo-center of $\alpha$ UMi AP}

\begin{tabular}{llrrrrrrrrrrrr}\hline\\[-1.5ex]
&&& \multicolumn{4}{c}{Prograde orbit} &&&
\multicolumn{4}{c}{Retrograde orbit}\\
 & & &&&&&&&\multicolumn{4}{c}{(Preferred
solution)}\\[0.5ex]\hline\\[-1.5ex]
Quantity [mas/year]& System && $\mu_{\alpha\ast}$ & m.e. & $\mu_\delta$ &
m.e. &&& $\mu_{\alpha\ast}$ & m.e. & $\mu_\delta$ &
m.e.\\[0.5ex]\hline\\[-1.5ex]
$\mu_{\rm{FK5}}$ & FK5 && + 38.30 & 0.23 & -- 15.20 & 0.35 &&&
+ 38.30 & 0.23 & -- 15.20 & 0.35\\
systematic correction &&& + 3.20 & 0.94 & -- 1.53 & 0.66 &&&
+ 3.20 & 0.94 & -- 1.53 & 0.66\\
$\mu_{\rm{FK5}}$ & HIP && + 41.50 & 0.97 & -- 16.73 & 0.75 &&&
+ 41.50 & 0.97 & -- 16.73 & 0.75\\[1.5ex]
$\mu_0$ & HIP && + 41.05 & 0.58 & -- 15.05 & 0.45 &&&
+ 40.56 & 0.58 & -- 14.67 & 0.45\\[1.5ex]
$\mu_{\rm{FK5}} - \mu_0$ & HIP &&  + 0.45 & 1.13 & -- 1.68 & 0.87 &&&
 + 0.94 & 1.13 & -- 2.06 & 0.87\\[1ex]\hline\\[-1.5ex]
$\mu_{\rm m}$ & HIP && + 41.17 & 0.50 & -- 15.49 & 0.39 &&&
+ 40.81 & 0.50 & -- 15.22 & 0.39\\[0.5ex]\hline\\[-1.5ex]
$\mu_{\rm H}$ & HIP && + 44.22 & 0.47 & -- 11.74 & 0.55 &&&
+ 44.22 & 0.47 & -- 11.74 & 0.55\\[0.5ex]\hline\\[-1.5ex]
$\Delta\mu =
\mu_{\rm H} - \mu_{\rm m}$ & HIP && + 3.05 & 0.69 & + 3.75 & 0.67 &&&
+ 3.41 & 0.69 & + 3.48 & 0.67\\[0.5ex]\hline
\end{tabular}

\end{table*}


\subsection{$\alpha$ UMi C and $\alpha$ UMi D}

In 1884 and 1890, Burnham (1894) measured two faint stars in the neighbourhood
of $\alpha$ UMi AB. In 1890.79, the component C had a separation of
44\,.$\!\!$"68 from A, and the component D 82\,.$\!\!$"83. According to the WDS
Catalogue, the apparent magnitudes of C and D are 13\,.$\!\!^m$1 and
12\,.$\!\!^m$1.

The nature of the components C and D is unclear.
The probability to find by chance a field star of the corresponding
magnitude with the observed separation around $\alpha$ UMi A
(galactic latitude $b = +26.\!\!^\circ46$ (Wielen 1974))
is of the order of 10 percent for each component.
This favours on statistical grounds a physical relationship
of the components C and D with A.
If C and D are physical members
of the Polaris system (instead of being optical components), their absolute
magnitudes in V would be +\,7\,.$\!\!^m$5 and +\,6\,.$\!\!^m$5. Due to the
low age of the Polaris system of about 70 million years (deduced from the
Cepheid $\alpha$ UMi A),
they would either just have reached the zero-age main
sequence, or they may still be slightly above this sequence (i.e.
pre-main-sequence objects, Fernie 1966).

\section{Astrometric orbit of $\alpha$ UMi AP}

In this section we determine the astrometric orbit of the photo-center of the
pair $\alpha$ UMi AP (i.e. essentially of A) with respect to the center-of-mass
of AP. We adopt all the elements of the spectroscopic orbit of A in the system
AP, derived by Kamper (1996). The remaining elements, i.e. the orbital
inclination $i$ and the nodal length $\Omega$, are basically obtained
from the following considerations:

The observed difference $\Delta\mu$ between the instantaneous proper motion of
$\alpha$ UMi A, provided by HIPPARCOS for an epoch
$T_{\rm{c, H}} \sim 1991.31$, and
the mean proper motion of $\alpha$ UMi A, provided by long-term, ground-based
observations, summarized in the FK5, is equal to the tangential component of
the orbital velocity of A with respect to the center-of-mass of the pair AP.
Using the spectroscopic orbit of A and the HIPPARCOS parallax, we can predict
$\Delta\mu$ for various adopted values of $i$ and $\Omega$. Comparing the
predicted values of $\Delta\mu$ with the observed difference $\Delta\mu$, we
find $i$ and $\Omega$. The length of the two-dimensional vector of $\Delta\mu$
gives us the inclination $i$; the direction of $\Delta\mu$ fixes then the
ascending node $\Omega$. Unfortunately, two values of $i$, namely $i$ and
180$^{\circ} - i$, predict the same value for $\Delta\mu$ (see Fig. 1).
This ambiguity corresponds to the fact that
$\Delta\mu$ itself does not allow us to differentiate between a prograde orbit
and a retrograde one. In the case of
$\alpha$ UMi AP, it is fortunate that the
ground-based observations of the Allegheny Observatory strongly favour
the retrograde orbit over the prograde one.


\begin{table*}

\caption[]{Proper-motion difference $\Delta\mu$ between
$\mu_{\rm H}$ and $\mu_{\rm m}$ of
the photo-center of $\alpha$ UMi AP at the epoch $T_{\rm{c, H}} = 1991.31$}

\begin{tabular}{lcrrrrrrrrrrrr}\hline\\[-1.5ex]
&&&& \multicolumn{4}{c}{Prograde orbit} &&& \multicolumn{4}{c}{Retrograde
orbit}\\
& & & &&&&&&&\multicolumn{4}{c}{(Preferred solution)}\\[0.5ex]\hline\\[-1.5ex]
Quantity & Unit &&& \multicolumn{2}{c}{Observed} &
\multicolumn{2}{c}{Predicted} &&&
\multicolumn{2}{c}{Observed} & \multicolumn{2}{c}{Predicted}
\\[0.5ex]\hline\\[-1.5ex]
$\Delta\mu_{\alpha\ast}$ & mas/year &&& + 3.05 & $\pm$ 0.69 & + 3.05 &&
& & + 3.41 & $\pm$ 0.69  & + 3.41 & \\
$\Delta\mu_\delta$ & mas/year &&& + 3.75 & $\pm$ 0.67 & + 3.75 &&&
& + 3.48 & $\pm$ 0.67  & + 3.48 & \\[1.5ex]
$\Delta\mu_{\rm{tot}}$ & mas/year &&& 4.83 & $\pm$ 0.61 & 4.83 & $\pm$ 0.32 &
& & 4.87 & $\pm$ 0.61 &  4.88 & $\pm$ 0.32\\
$\Theta_{\Delta\mu}$ & $^\circ$ &&& 39.1\hphantom{0} & $\pm$ 8.8\hphantom{0} &
39.1\hphantom{0} & $\pm$ 3.5\hphantom{0} &&& 44.4\hphantom{0}
& $\pm$ 8.8\hphantom{0} &
44.4\hphantom{0} & $\pm$ 3.4\hphantom{0}\\[0.5ex]\hline
\end{tabular}

\end{table*}


\subsection{The determination of $\Delta\mu$}

\subsubsection*{{\it 3.1.1. The proper motion of the center-of-mass of AP}}

We determine first the proper motion $\mu_{\rm{cms}(AP)}$ of the center-of-mass
of the pair AP. ($\mu$ is used here for $\mu _{\alpha\ast} = \mu_\alpha
\cos\delta$ or for $\mu_\delta$). The proper motion $\mu_{\rm{FK5}}$ of
$\alpha$ UMi given in the FK5 should be very close to
$\mu_{{\rm cms (AP)}}$, since
the ground-based data are averaged in the
FK5 over about two centuries, which is much larger than the
orbital period of AP of about 30 years. In Table 1, we list $\mu_{\rm{FK5}}$ in
the FK5 system and, by applying appropriate  systematic corrections, in the
HIPPARCOS/ICRS system. The mean errors of $\mu_{\rm{FK5}}$ in the HIPPARCOS
system include both the random error of $\mu_{\rm{FK5}}$ and the uncertainty
of the systematic corrections.

Another determination of $\mu_{\rm{cms} (AP)}$
is based on the positions $x_{\rm H}
(T_{\rm{c, H}})$ and $x_{\rm{FK5}} (T_{\rm{c, FK5}})$
at the central epochs
$T_{\rm{c, H}}$ and $T_{\rm{c, FK5}}$ of
the HIPPARCOS Catalogue and of the FK5. The designation $x$ stands for
$\alpha_\ast = \alpha \cos \delta$ or $\delta$, where $\alpha$ is the right
ascension and $\delta$ the declination of $\alpha$ UMi. The position
$x_{\rm{FK5}} (T_{\rm{c, FK5}})$
represents a time-averaged, `mean' position in
the sense of Wielen (1997). Before being used, $x_{\rm{FK5}}
(T_{\rm{c, FK5}})$ must be reduced to the HIPPARCOS/ICRS system.

The HIPPARCOS position is (approximately) an `instantaneously' measured
position of the photo-center of AP. Before combining the HIPPARCOS position
with $x_{\rm{FK5}}$ to a mean proper motion $\mu_0$,
we have to reduce $x_{\rm H}$ to
the mean position
$x_{\rm{mean \, ph (AP), H}} (T_{\rm{c, H}})$ of the photo-center of AP
at time $T_{\rm{c, H}}$. This is
done by going first from $x_{\rm H} (T_{\rm{c, H}})$ to the center-of-mass
$x_{\rm{cms(AP), H}} (T_{\rm{c, H}})$
by subtracting from $x_{\rm H}$ the orbital
displacement
$\Delta x_{\rm{orb, ph (AP)}} (T_{\rm{c, H}})$
predicted by the astrometric orbit of
the photo-center of AP. Then we have to add to
$x_{\rm{cms(AP), H}} (T_{\rm{c, H}})$ the
(constant) off-set between the mean position of the photo-center
$x_{\rm{mean} \, ph (AP)}$ and the center-of-mass (see Fig. 1). Using now
$\alpha_\ast$ and $\delta$, we obtain
\begin{eqnarray}
\lefteqn{
\alpha_{\ast, {\rm{mean \, ph (AP), H}}} (T_{\rm{c, H}})
= \alpha_{\ast,{\rm{cms(AP), H}}} (T_{\rm{c, H}})}\nonumber\\
& & \hspace*{0.5cm}
- \,  \frac{3}{2} \, e \, a_{{\rm ph (AP)}} (\cos \omega \sin
\Omega + \sin \omega \cos \Omega \cos i) \,\, ,\\                      
\lefteqn{
\delta_{\rm{mean \, ph (AP), H}} (T_{\rm{c, H}})
= \delta_{\rm{cms(AP), H}} (T_{\rm{c, H}})}\nonumber\\
& & \hspace*{0.5cm}
- \,  \frac{3}{2} \, e \, a_{{\rm ph (AP)}} (\cos \omega \cos
\Omega - \sin \omega \sin \Omega \cos i) \,\, ,                         
\end{eqnarray}
where $a_{{\rm ph (AP)}}$
is the semi-major axis of the orbit of the photo-center of
AP around the center-of-mass of AP. The other elements of this orbit are:
eccentricity $e$, inclination $i$, longitude of periastron $\omega$, position
angle of the ascending node $\Omega$, orbital period $P$, epoch of periastron
passage $T_{\rm{peri}}$. The quantities in the Eqs. (1) and (2) which follow
after $- \, \frac{3}{2} \, e$ are just the
Thiele-Innes elements $B$ and $A$. The equations use the fact that,
in the orbital plane, the time-averaged position is located on the major axis,
towards the apastron, at a distance of $\frac{3}{2} \, e \, a$ from the
center-of-mass.

If $\alpha_\ast$ and $\delta$ would change linearly with time, we could
determine the mean proper motion $\mu_0$ from
\begin{equation}
\mu_0 = \frac{x_{\rm{mean \, ph (AP), H}} (T_{\rm{c, H}}) - x_{\rm{FK5}}
(T_{\rm{c, FK5}})} {T_{\rm{c, H}} - T_{\rm{c, FK5}}} \,\, .          
\end{equation}
However, for Polaris we should use more accurate formulae because it is so
close to the celestial pole. We determine $\mu_0$ strictly by
requiring that $\mu_0
(T_{\rm{c, H}})$ is that proper motion which brings the object from
$x_{\rm{mean \, ph (AP), H}} (T_{\rm{c, H}})$ to
$x_{\rm{FK5}} (T_{\rm{c, FK5}})$. For
calculating
the (small) foreshortening effect, we have adopted the radial velocity of the
center-of-mass of AP, $v_{\rm r} = \gamma = - 16.42$ km/s (Kamper 1996).

The agreement between the two mean proper motions $\mu_{\rm{FK5}}$ and
$\mu_0$ is rather good (Table 1). For determining the best value
$\mu_{\rm m}$ of the mean motion of the photo-center,
which is equal to the proper motion of the center-of-mass, we take the weighted
average of $\mu_{\rm{FK5}}$ and $\mu_0$. Since the orbital corrections to
$x_{\rm H}$
are different for the prograde and retrograde orbits, we have two values for
$\mu_0$ and hence for $\mu_{\rm m}$. In both cases, we had to iterate the
determinations of the orbital elements ($i$ and $\Omega$) and of $\mu_0$ (and
hence $\mu_{\rm m}$), since $\mu_0$ depends on the orbital corrections.
The values for
$\mu_{\rm m} = \mu_{\rm{cms}}$ finally adopted are listed in Table 1.

\subsubsection*{{\it 3.1.2. The HIPPARCOS proper motion $\mu_{\rm H}$}}

The HIPPARCOS proper motion $\mu_{\rm H}$ of Polaris (ESA 1997) refers to the
photo-center of AP. Basically, $\mu_{\rm H}$ is the sum of the proper motion
$\mu_{\rm{cms} (AP)}$ of the center-of-mass (cms) of AP and of the orbital
motion $\Delta\mu_{\rm{orb}, ph (AP)}$ (abbreviated as $\Delta\mu$) of the
photo-center of AP with respect to the cms of AP at time $T_{\rm{c, H}}$:
\begin{equation}
\mu_{\rm H} (T_{\rm{c, H}})
= \mu_{\rm{cms} (AP)} + \Delta\mu (T_{\rm{c, H}}) \,\, .   
\end{equation}
During the reduction of the HIPPARCOS data, a linear `standard' solution was
applied to Polaris. The variation of $\Delta\mu (t)$ during the period of
observations of about three years was neglected. This slightly complicates the
comparison of the observed $\Delta\mu$ with the orbital ephemerides. In Sect.
3.1.5 we assume that $\mu_{\rm H}$ is obtained from a linear fit to
quasi-continuously measured true positions
over a time interval $D_{\rm H}$, centered at time
$T_{\rm{c, H}}$. From the correlation
coefficients given in the HIPPARCOS Catalogue, we derive for the central epochs
$T_{\alpha, {\rm H}} = 1991.26$
and $T_{\delta, {\rm H}} = 1991.35$. We neglect the slight
difference between
$T_{\alpha, {\rm H}}$ and $T_{\delta, {\rm H}}$ and use the average of
both, namely $T_{\rm{c, H}} = 1991.31$. From the epochs of the individual
observations of Polaris by HIPPARCOS, we estimate $D_{\rm H} = 3.10$ years.

\subsubsection*{{\it 3.1.3. The observed value of $\Delta\mu$}}

The observed value of $\Delta\mu$ is derived from
\begin{equation}
\Delta\mu (T_{\rm{c, H}})
= \mu_{\rm H} (T_{\rm{c, H}}) - \mu_{\rm{cms} (AP)} (T_{\rm{c, H}})
                                                            \,\, .     
\end{equation}

The values of $\Delta\mu$ in $\alpha_\ast$ and $\delta$, derived from Eq. (5),
are listed in Tables 1 and 2. Table 2 gives also the total length
$\Delta\mu_{\rm{tot}}$ of the $\Delta\mu$ vector,
\begin{equation}
(\Delta\mu_{\rm{tot}})^2 = (\Delta\mu_{\alpha\ast})^2 + (\Delta\mu_\delta)^2
                                                       \,\, ,          
\end{equation}
and the position angle $\Theta_{\Delta\mu}$ of the $\Delta\mu$ vector,
\begin{equation}
\Theta_{\Delta\mu} = \arctan (\Delta\mu_{\alpha\ast} / \Delta\mu_\delta) \,\, .
\end{equation}                                                         
All the values are valid for the equinox J2000 in the HIPPARCOS/ICRS system
at the epoch $T_{\rm{c, H}} = 1991.31$.

Since $\mu_{\rm{cms}}$ depends on the direction of motion in the orbit
(prograde or retrograde), this is also
true for $\Delta\mu$, and we obtain therefore two values for
$\Delta\mu$. Table 2 shows that $\Delta\mu_{\rm{tot}} \sim 5$ mas/year and
$\Theta_{\Delta\mu}$ are statistically quite significant and rather well
determined. A value of $\Delta\mu_{\rm{tot}} = 4.87$ mas/year corresponds to a
tangential velocity of 3.05 km/s. Hence the `instantaneous' HIPPARCOS proper
motion of Polaris has a significant `cosmic error' (Wielen 1995a,\,b,
1997, Wielen et al. 1997, 1998, 1999a,\,b) with respect to the motion of the
center-of-mass.
If Polaris were not already known as a close binary, our $\Delta\mu$
method (Wielen et al. 1999a) would have detected Polaris to be
a $\Delta\mu$ binary because of its large test parameter
$F_{\rm{FH}} =$ 6.18 for $\mu_{\rm{FK5}} - \mu_{\rm H}$.

\subsubsection*{{\it 3.1.4. The photo-center of $\alpha$ UMi AP}}

The HIPPARCOS observations refer to the photo-center of $\alpha$ UMi AP, since
the pair is not resolved by HIPPARCOS.
The `phase' used in constructing the HIPPARCOS
Catalogue is practically identical to the phase of the photo-center,
because the magnitude difference $\Delta m_{{\rm AP}}$ of more than 6 mag
between A and P is quite large
and because the separation between A and P at $T_{\rm H}$
was rather moderate (about 93 mas).
It can also be shown that the component B does not significantly affect
the HIPPARCOS measurements of AP, because of
$\Delta m_{\rm{V, B-AP}} = $ 6.61,
in spite of its separation $\rho = 18"$.
The
HIPPARCOS observations have been carried out in a broad photometric band called
{\rm Hp}. The photo-center refers therefore to this photometric system.

The spectroscopic orbit, however, refers to component A. We have therefore to
transform the value
$a_{\rm A} \sin i$ of the spectroscopic orbit into $a_{{\rm ph (AP)}}
\sin i$ for obtaining an astrometric orbit of the photo-center of AP. The
relation between $a_{\rm A}$ and $a_{{\rm ph (AP)}}$ is given by
\begin{equation}
a_{{\rm ph (AP)}} = (1 - \frac{\beta}{B}) \, a_{\rm A}  \,\, ,
\end{equation}
where $B$ and $\beta$ are the fractions of the mass ${\cal M}$ and the
luminosity $L$ of the secondary component P:
\begin{eqnarray}
B & = & \frac{{\cal M}_{\rm P}}
{{\cal M}_{\rm A} + {\cal M}_{\rm P}}  \,\, , \\[1ex]      
\beta & = & \frac{L_{\rm P}}{L_{\rm A} + L_{\rm P}}
= \frac{1}{1 + 10^{0.4\,\Delta m_{{\rm AP}}}} \,\,.
\end{eqnarray}                                                          
$\Delta m_{{\rm AP}}$ is the magnitude difference between A and P:
\begin{equation}
\Delta m_{{\rm AP}} = m_{\rm P} - m_{\rm A} = M_{\rm P} - M_{\rm A} \,\, . 
\end{equation}
Using the results given in Table 5, we find for $\alpha$ UMi AP
\begin{equation}
1 - \frac{\beta}{B} = 0.988 \,\, ,                                
\end{equation}
with an estimated error of about $\pm 0.010$. For calculating $\beta$, we have
assumed that $\Delta m_{{\rm AP}}$ is the same in {\rm Hp} as in  V. This
approximation is fully justified for our purpose. We derive (see the end of
Sect. 3.1.6) from Kamper (1996) for component A:
\begin{equation}
a_{\rm A} \sin i
= 2.934 \pm 0.028 \, {\rm AU} \,\, .                       
\end{equation}
The HIPPARCOS parallax of Polaris (ESA 1997) is
\begin{equation}
p_{\rm H}
= 7.56 \pm 0.48 \, {\rm mas} \,\, .                              
\end{equation}
This leads to
\begin{equation}
a_{\rm A} \sin i
= 22.18 \pm 1.42 \, {\rm mas} \,\, .                         
\end{equation}
Using Eqs. (8), (12), and (15), we obtain for the photo-center
\begin{equation}
a_{{\rm ph (AP)}} \sin i
= 21.91 \pm 1.42 \, {\rm mas} \,\, .                  
\end{equation}

\subsubsection*{{\rm 3.1.5. The predicted value of $\Delta\mu$}}

For predicting $\Delta\mu$, we use four elements $(P, e, T_{\rm{peri}},
\omega)$ of the spectroscopic orbit derived by Kamper (1996),
and $a_{{\rm ph (AP)}}
\sin i$ according to Eq. (16), all listed in Table 4 . In addition we adopt
various values of $i$ and $\Omega$ in order to produce predicted
values of $\Delta\mu$ at
time $T_{\rm{c, H}}$ as a function of $i$ and $\Omega$.

Since the observed value of $\Delta\mu$ is not an instantaneously measured
tangential velocity, we mimic the HIPPARCOS procedure of determining
$\mu_{\rm H}$.
We calculate the positions $\Delta x_{\rm{orbit, ph (AP)}}(t) \equiv \Delta x
(t)$ of the photo-center of AP with respect to the cms of AP as a function of
time, using standard programs for the ephemerides of double stars. We then
carry out a linear least-square fit to these positions over a time interval of
length $D_{\rm H} = 3.10$ years, centered at $T_{\rm{c, H}} = 1991.31$:
\begin{eqnarray}
\Delta x_{{\rm av}} (T_{\rm{c, H}})
& = & \frac{1}{D_{\rm H}} \, \int\limits^{+D_{\rm H}/2}_{-D_{\rm H}/2}
\,
\Delta x (T_{\rm{c, H}} + \tau) d \tau  \,\, ,                 \\[1ex] 
\Delta\mu_{{\rm av}} (T_{\rm{c, H}})
& = & \frac{12}{D^3_{\rm H}} \,
\int\limits^{+D_{\rm H}/2}_{-D_{\rm H}/2}\,
\Delta x (T_{\rm{c, H}} + \tau) \tau \, d \tau  \,\, .                 
\end{eqnarray}
Tests have shown that especially
$\Delta\mu_{{\rm av}}$ is not very sensitive against
small changes in the slightly uncertain quantity $D_{\rm H}$.
Actual numbers for the
predicted values
$\Delta\mu_{{\rm av}} (T_{\rm{c, H}})$ are given in the Tables 2 and 3.

\subsubsection*{{\it 3.1.6. The problem of $T_{\rm{peri}}$ and of
$a_{\rm A} \sin i$}}

In his paper, Kamper (1996, his Table III) gives for his best orbit (DDO+Lick
Data) a value for $T_{\rm{peri}} = 1928.48 \pm 0.08$. This is exactly the
value derived by Roemer (1965) from the Lick Data and also quoted in Kamper's
Table III under `Lick Data'. There are three possibilities for this
coincidence: (1) Kamper has adopted this value of $T_{\rm{peri}}$ as a fixed
input value from Roemer. Nothing is said about this in his paper. (2) Kamper
found from a full least-square solution by chance the same values for
$T_{\rm{peri}}$ and its mean error as quoted for Roemer. Such a mere accident
is highly improbable. (3) The identical values of $T_{\rm{peri}}$ and its mean
error in the two columns of Kamper's Table III occured due to a mistake or
misprint. However, Kamper has not published any erratum in this direction.

Dr. Karl W. Kamper died in 1998 (Bolton 1998). We tried to get clarification on
the problem of $T_{\rm{peri}}$ from colleagues of Dr. Kamper, but they were
unfortunately unable to help us in this respect. Hence we are inclined to
accept the possibility (1). However, even then there is an additional problem
with the mean error of $T_{\rm{peri}}$. Kamper has obviously overlooked that
Roemer (1965) gave {\it probable} errors instead of {\it mean} errors. Hence
the mean error of $T_{\rm{peri}}$ according to Roemer should read $\pm$\,0.12
in Kamper's Table III.

For our purpose, a value of $T_{\rm{peri}}$ closer to
$T_{\rm{c, H}}$ should be chosen.
Using $T_{\rm{peri}} = 1928.48 \pm 0.12$ and $P =
29.59 \pm 0.02$ years, we obtain an alternative value (two periods later) of
\begin{equation}
T_{\rm{peri}} = 1987.66 \pm 0.13 \,\, .                                
\end{equation}
We have tested this value by carrying out an unweighted least-square fit to the
mean radial velocities of $\alpha$ UMi A listed in Table II of Kamper (1996).
In this solution we solved for $T_{\rm{peri}}$ only, while we adopted all the
other spectroscopic elements as given by Kamper (1996). We obtained
$T_{\rm{peri}} = 1987.63 \pm 0.25$, in good agreement with Eq. (19). However,
the formally most accurate radial velocity listed in the last line of Kamper's
Table II does not fit perfectly (O--C\,=\,--\,0.15 km/s) his final orbit with
$T_{\rm{peri}}$ according to Eq. (19), but rather indicates the
value of $T_{\rm{peri}} = 1987.27$. The independent
radial-velocity data published by
Dinshaw et al. (1989) lead us to $T_{\rm{peri}} = 1987.57$ with a very small
formal error. This is in good agreement with Eq. (19). Hence we have finally
adopted $T_{\rm{peri}}$ as
given by Eq. (19). An error of $\pm$\,0.13 years introduces
errors of
$\pm 0\,.\!\!^\circ3$ in $i$ and of $\pm 2\,.\!\!^\circ 0$ in $\Omega$,
which are small compared
to the errors in $i$ and $\Omega$ due to the uncertainties in
$a_{{\rm ph (AP)}} \sin i$ and in the observed value of $\Delta\mu$.

The values of $a_{\rm A} \sin i, K_{\rm A}, P$,
and $e$ given by Kamper (1996) in
his Table
III under DDO+Lick Data are unfortunately not consistent.
If we accept $K_{\rm A},
P$, and $e$, we find $a_{\rm A} \sin i = 2.934$ AU, while Kamper gives in his
Table III 2.90 AU. In the {\it text} of his paper, Kamper gives 2.9 AU for
$a_{\rm A}
\sin i$. Has he rounded 2.934 to 2.9 and later inserted this rounded value
as 2.90 into his Table III ?
We prefer to trust $K_{\rm A}$ and $e$, and hence we
use for $a_{\rm A} \sin i$ the value of 2.934 AU (see Sect. 3.1.4) in our
investigation.


\begin{table}

\caption[]{Determination of the inclination $i$ from $\Delta\mu_{\rm{tot}}$}

\tabcolsep2.2mm
\begin{tabular}{crrrcrr}\hline\\[-1.5ex]
\multicolumn{3}{c}{Prograde orbit} && \multicolumn{3}{c}{Retrograde orbit}\\
&&&& \multicolumn{3}{c}{(Preferred solution)}\\[0.5ex]\hline\\[-1.5ex]
& \multicolumn{2}{c}{$\Delta\mu_{\rm{tot}}$} &&&
\multicolumn{2}{c}{$\Delta\mu_{\rm{tot}}$}\\
& \multicolumn{2}{c}{[mas/years]} &&& \multicolumn{2}{c}{[mas/years]}
\\[0.5ex]\hline\\[-1.5ex]
Observed: & 4.83 & $\pm$ 0.61 && Observed: & 4.87 & $\pm$
0.61\\[0.5ex]\hline\\[-1.5ex]
Predicted &&&& Predicted & & \\[0.5ex]
for $i$ [$^\circ$]: &&& & for $i$ [$^\circ$]:& &\\[0.5ex]
30 & 9.32 &&& 150 & 9.32 &\\
40 & 6.58 &&& 140 & 6.58 &\\
45 & 5.63 &&& 135 & 5.63 &\\
50 & 4.85 &&& 130 & 4.85 &\\
55 & 4.19 &&& 125 & 4.19 &\\
60 & 3.64 &&& 120 & 3.64 &\\
70 & 2.78 &&& 110 & 2.78 &\\[1.5ex]
\hphantom{.0}50.1 & 4.83 & $\pm$ 0.32 &&
\hphantom{O}130.2 & 4.88 & $\pm$ 0.32 \\[0.5ex]\hline\\
\end{tabular}

\end{table}


\subsection{The astrometric orbit}

\subsubsection*{{\it 3.2.1. Determination of the inclination $i$}}

In Table 3 we compare the {\it observed} values of $\Delta\mu_{\rm{tot}}$ (from
Sect. 3.1.3 and Table 2) with the {\it predicted} values of
$\Delta\mu_{\rm{tot}}$ (using the
procedures described in Sect. 3.1.5 and the elements $P, e,
T_{\rm{peri}}, \omega$ and $a_{{\rm ph (AP)}} \sin i$
given in Table 4) for different
trial values of the inclination $i$. The length $\Delta\mu_{\rm{tot}}$ of the
vector $\Delta\mu$ is obviously not a function of the nodal direction $\Omega$.
The mean error of the predicted value of $\Delta\mu_{\rm{tot}}$ includes the
uncertainties in all the orbital elements except in $i$ and $\Omega$.

The best agreement between the observed and predicted values of
$\Delta\mu_{\rm{tot}}$ occurs for $i = 50\,.\!\!^\circ1$ (prograde orbit) and
for $i = 130\,.\!\!^\circ2$ (retrograde orbit).
The uncertainties in the observed value of
$\Delta\mu_{\rm{tot}}$ and in the orbital elements
(mainly in $a_{{\rm ph (AP)}}
\sin i$) lead to an uncertainty in $i$ of $\pm 4\,.\!\!^\circ8$.

Our values for the inclination $i$ of the two orbits do not fulfill strictly
the expected relation $i_{\rm{retrograde}} = 180^\circ - i_{\rm{prograde}}$.
The reason is the following: $i$ is determined (Table 3) from two slightly
different values $\Delta\mu$ for the prograde and retrograde orbits (Table 2).
The difference in the $\Delta\mu$ values stems from a slight difference in
$\mu_0$ and hence in $\mu_{\rm m}$ (Table 1), and this difference in $\mu_0$ is
caused by a difference in
$x_{\rm{mean \, ph (AP), H}}$ (Eq. (3)). The difference
in the position
$x_{\rm{mean \,ph (AP), H}}$ of the mean photo-center is due to
the small, but totally different corrections which have to be added to the
observed HIPPARCOS position
$x_{\rm{ph (AP), av, H}}$ in order to obtain the mean
photo-center (see Table 6 and Fig. 1). As mentioned already  at the end of
Sect. 3.1.1., we had to iterate our procedure of determining $i$ and
$\Omega$, since the corrections depend on these orbital elements.

The fit between the observed and predicted values of $\Delta\mu$ is rather
pleasing. It is not granted that such a fit is always possible.
In the case of Polaris, for example, the spectroscopic orbit and the HIPPARCOS
parallax
together require a minimum value of $\Delta\mu_{\rm{tot}}$ of 2.00 mas/year,
which occurs for $i = 90^\circ$.
There is no formal upper limit for $\Delta\mu_{\rm{tot}}$ for $i
\rightarrow 0$. However, the requirement that the component P is not visible in
the combined spectrum of AP gave for a main-sequence companion P a spectral
type later than A8V (Sect.
2.2), or ${\cal M}_{\rm P} < 1.8\,{\cal M}_\odot$. Combined
with the mass function of the spectroscopic orbit,
$f ({\cal M}) = ({\cal M}_{\rm P}
\sin i)^3/({\cal M}_{\rm A} + {\cal M}_{\rm P})^2 = 0.02885 {\cal M}_\odot$,
and with a
reasonable estimate of
${\cal M}_{\rm A} ({\cal M}_{\rm A} > 5\,{\cal M}_\odot)$,
this
gives a lower limit for $i$ of about $i > 37^\circ$, which
corresponds to $\Delta\mu_{\rm{tot}} < 7$ mas/year.
Our observed value of $\Delta\mu_{\rm{tot}}$ of about 5 mas/year fulfills
nicely the range condition of 2 mas/year $< \Delta\mu_{\rm{tot}} < 7$
mas/year.

\subsubsection*{{\it 3.2.2. Determination of the nodal length $\Omega$}}

Having fixed the inclination $i$ in Sect. 3.2.1, we now determine $\Omega$
from a comparison of the observed and predicted values of the direction
$\Theta_{\Delta\mu}$ of the vector $\Delta\mu$. The difference (modulo
360$^\circ$) between the observed value of $\Theta_{\Delta\mu}$ and the
predicted value of $\Theta_{\Delta\mu}$ for $\Omega = 0$ gives just that
desired value of $\Omega$ for which the observed and predicted values of
$\Theta_{\Delta\mu}$ agree. We find $\Omega = 276\,.\!\!^\circ2$ for the
prograde orbit and $\Omega = 167\,.\!\!^\circ1$ for the retrograde orbit.
The uncertainties in the
observed value of $\Theta_{\Delta\mu}$ and in the orbital elements (now mainly
in $i$ and $T_{\rm{peri}}$) lead to an
uncertainty in $\Omega$ of $\pm 9\,.\!\!^\circ5$ or $\pm 9\,.\!\!^\circ4$.

The quality of the fit in the components of $\Delta\mu$ in $\alpha_\ast$ and
$\delta$ can be judged from the data given in Table 2. The overall agreement is
quite good.

\subsubsection*{{\it 3.2.3. The ambiguity problem of $i$}}

If we know only the vector $\Delta\mu$ at one epoch and the spectroscopic orbit
of a binary, then there is an ambiguity ($i$ or 180$^\circ - i$) in the
inclination $i$, i.e. in
the direction of motion in the astrometric orbit. In the prograde (or `direct')
orbit ($i < 90^\circ$), the position angle of P relative to A increases with
time, in the
retrograde orbit ($i > 90^\circ$) it decreases. The reason for the ambiguity is
the fact that
$\Delta\mu$ itself does not indicate whether the orbit will turn to the
left-hand side or to the right-hand side (see Fig. 1).

In principle, the knowledge of the mean position of the photo-center predicted
by the FK5 for
$T_{\rm{c, H}} = 1991.31$ would resolve the ambiguity. However, the
mean errors of this predicted position of $\pm$\,72 mas in $\alpha_\ast$ and
$\pm$\,66 mas in $\delta$ are so large with respect to the differences between
$x_{\rm H} (T_{\rm{c, H}})$ and
$x_{\rm{mean \, ph} (AP)} (T_{\rm{c, H}})$, which are
less than 26 mas (Table 6), that this method is not useful in our case.

At present, the best solution of the ambiguity problem is provided by the
results of the photographic observations carried out at the Allegheny
Observatory, which we discussed already in Sect. 2.2\,. While the full
astrometric orbit based on the Allegheny data (Kamper 1996) is not very
trustworthy, the Allegheny data give strong preference for a retrograde orbit
(in contrast to a direct one). This can be seen best in Fig. 3 of Kamper
(1996): The minimum of the residuals (dashed line) occurs for $i > 120^\circ
\, (\cos i < - 0.5)$, and for this range of $i$ the semi-major axis derived
from the Allegheny data is quite reasonable. For our preferred value of
$i\,(130\,.\!\!^\circ2)$, we
read off from Kamper's Fig. 3 a value of $a_{\rm{ph (AP)}}
\sim 28$ mas with an estimated uncertainty of $\pm$\,9 mas. This is in very
good agreement with our result, 28.7\,$\pm$\,2.8 mas. Even the nodal length
$\Omega$ derived by Kamper $(175^\circ)$ is compatible with our
result $(167^\circ \pm 9^\circ)$. Kamper's determination of $i\,( 179^\circ)$
is very uncertain and therefore not in contradiction to our value
(130$^\circ$). He himself says in the text of the paper that `all inclinations
between 135$^\circ$ and 180$^\circ$ are equally satisfactory' in fitting the
Allegheny data. (There is a small mistake in Kamper's discussion of this point:
He claims in the text `that the minimum scatter is for an inclination of almost
90$^\circ$, which results in a face-on orbit'. The relative clause after
90$^\circ$, his own Fig. 3 and his Table III all indicate that `90$^\circ$'
should be replaced by `180$^\circ$'.)

A new astrometric space mission will immediately resolve the ambiguity, since
it shall then be clear to which side of our $\Delta\mu$ vector (Fig. 1) the
orbit will have turned over. Probably the much higher accuracy of a new space
mission will allow to determine the direction (and amount) of the
instantaneous acceleration (i.e. to obtain a `G solution' in the HIPPARCOS
terminology, if not even a full orbital `O solution').


\begin{table}

\caption[]{Orbital elements of $\alpha$ UMi AP}

\begin{tabular}{lrr}\hline\\[-1.5ex]
Quantity & Prograde orbit & Retrograde orbit\hphantom{R}\\
& &                         (Preferred solution)\\[0.5ex]\hline\\[-1.5ex]
$v_{\rm r} = \gamma$ [km/s] & \multicolumn{2}{c}
{-- 16.42 $\pm$ 0.03 \hphantom{0}}\\
$K_{\rm A}$ [km/s] & \multicolumn{2}{c}
{ \hphantom{.} 3.72 $\pm$ 0.03}\\
$a_{\rm A} \sin i$ [AU] &  \multicolumn{2}{c}
{ \hphantom{.} 2.934 $\pm$ 0.028}\\
$a_{\rm A} \sin i$ [mas] & \multicolumn{2}{c}
{22.18 $\pm$ 1.42}\\
$a_{{\rm ph (AP)}} \sin i$ [mas] & \multicolumn{2}{c}
{21.91 $\pm$ 1.42}\\[0.5ex]\hline\\[-1.5ex]
$a_{{\rm ph (AP)}}$ [mas] & 28.56 $\pm$ 2.73 & 28.69 $\pm$ 2.75\\
$a_{\rm A} $ [mas] & 28.91 $\pm$ 2.74 & 29.04 $\pm$ 2.77\\[1.5ex]
$P$ [years] & \multicolumn{2}{c}
{ \hphantom{0} 29.59 $\pm$ 0.02 \hphantom{.} }\\
$e$         & \multicolumn{2}{c}
{ \hphantom{0} 0.608 $\pm$ 0.005}\\
$T_{\rm{peri}}$ & \multicolumn{2}{c}
{1987.66 $\pm$ 0.13 \hphantom{0} }\\
$\omega$ [$^\circ$] & \multicolumn{2}{c}
{ \hphantom{.} 303.01 $\pm$ 0.75 \hphantom{0} }\\[1.5ex]
$i$ [$^\circ$] & 50.1 $\pm$ 4.8 & 130.2 $\pm$ 4.8\\
$\Omega$ [$^\circ$] & 276.2 $\pm$ 9.5 & 167.1 $\pm$ 9.4\\[0.5ex]\hline
\end{tabular}

\end{table}


\subsubsection*{{\it 3.2.4. Resulting orbits of $\alpha$ UMi AP}}

The resulting orbits of the photo-center of $\alpha$ UMi AP are listed in Table
4. As explained in Sect. 3.2.3, the retrograde orbit should be preferred. The
semi-major axes of the orbits of $\alpha$ UMi A and P itself, relative to the
center-of-mass of AP, and that of P relative to A, are given in Table 5.

In Fig. 1, the two orbits (prograde and retrograde) of the photo-center of AP
are illustrated. The zero-point of the coordinates $\Delta\alpha_\ast$ and
$\delta$ is the HIPPARCOS position $x_{\rm H} (T_{\rm{c, H}})$ at epoch
$T_{\rm{c, H}} = 1991.31$.
The zero-point is then comoving with the center-of-mass (cms) of
either
the prograde orbit or the retrograde one. Therefore the orbits stay fixed in
these coordinates. Since the proper motion $\mu_{\rm{cms}}$ of the cms of the
two orbits differs slightly (Table 6), the linear motion of the position
$x_{\rm H} (t)$ predicted from the HIPPARCOS
Catalogue, differs slightly for the two cases. The
indicated motion of $x_{\rm H}$ corresponds to $\Delta\mu$ (Table 2). Hence by
construction, the motion of $x_{\rm H}$ is a tangent to the corresponding orbit,
except for the slight difference between the averaged position
and the instantaneous position of the photo-center at $T_{\rm{c, H}}$.
The dots on the orbits mark the positions in intervals of one year, the
years 1990, 1995, 2000, etc. being accentuated by a larger dot. We indicate
also the true major axis, on which periastron, center-of-mass, mean
photo-center, and apastron are located. In addition we plot the line of nodes.
The position of the ascending node is indicated by $\Omega$. Fig. 1
demonstrates clearly that the position predicted by HIPPARCOS is drifting away
from the actual position of the photo-center of AP.


\begin{table}

\caption[]{Physical properties of $\alpha$ UMi A and P}

\tabcolsep1.45mm
\begin{tabular}{lclll}\hline\\[-1.5ex]
Quantity & Units & Combined  & Component  & Component\\
& & \hphantom{AP}A+P & \hphantom{AAA}A & \hphantom{AAA}P\\
[0.5ex]\hline\\[-1.5ex]
$m_{\rm V}$ & [mag] & + 1.982 & + 1.985 & + 8.5 $\pm$ 0.4\\
$M_{\rm V}$ & [mag] & -- 3.63 $\pm$ 0.14 & -- 3.62 $\pm$ 0.14 & + 2.9 $\pm$
0.4\\[1.5ex]
${\cal M}$ & [${\cal M}_\odot$] & 7.54 $\pm$ 0.6 & 6.0 $\pm$ 0.5 & 1.54 $\pm$
0.25\\
Spec. type & & & F7-F8 Ib-II & F0V\\
Age $\tau$ & [years] & & 7\,$\cdot$\,10$^7$\\
a & [mas] & 142 $\pm$ 21 *)& 29.0 $\pm$ 2.8 c)& 113 $\pm$ 21 c)\\
a &  [AU] & 18.8 $\pm$ 2.8 *)& 3.84 $\pm$ 0.37 c)& 15.0 $\pm$ 2.8 c)\\
Mean sep. & [mas]  & \hphantom{00} 135 *)\\[0.5ex]\hline\\[-1ex]
\multicolumn{5}{l}{
\hspace{-0.10 cm}
*) orbit of P relative to A;
\hspace{0.10 cm}
c) orbit with respect to the cms of AP.}\\
\end{tabular}

\end{table}


\subsubsection*{{\it 3.2.5. Derived physical properties of $\alpha$ UMi P}}

In Table 5, we summarize some physical properties of the components A and P of
$\alpha$ UMi.

The mass of $\alpha$ UMi A is derived from the mass-luminosity relation for
Cepheids given by Becker et al. (1977). Since we use the luminosity based on
the HIPPARCOS distance (132 pc), our value of 6 ${\cal M}_\odot$ is higher than
that of other authors who have used a smaller distance.

The age of $\alpha$ UMi A, and therefore of the whole system of Polaris, can be
estimated from the period-age relation for Cepheids (Becker et al. 1977,
Tammann 1969). Using $P_0 = 5.64$ days (see Sect. 2.1), we derive an age $\tau$
of about $7 \cdot 10^7$ years.

The spectroscopic orbit provides the mass function $f ({\cal M}) = 0.02885 \,
{\cal M}_\odot$. Adopting the inclination $i = 130\,.\!\!^\circ2$ of the
retrograde orbit and ${\cal M}_{\rm A} = 6.0 \, {\cal M}_\odot$
for the Cepheid, we
obtain ${\cal M}_{\rm P} = 1.54 \, {\cal M}_\odot$ for the component P.
Using this
value for ${\cal M}_{\rm P}$,
we estimate for a star on the zero-age main sequence an absolute
magnitude of
$M_{\rm V} = + 2.9$ and a spectral type of F0V. The magnitude difference
$\Delta m_{{\rm V, AP}}$
between A and P is then about 6\,.$\!\!^m$5 . As mentioned
in Sect. 2.2, a White Dwarf
is ruled out by the IUE spectra and the low age of Polaris.
Our estimate for
${\cal M}_{\rm P}$ itself would not violate the Chandrasekhar limit
for White Dwarfs, if we consider the uncertainty in ${\cal M}_{\rm P}$ of
$\pm 0.25 \, {\cal M}_\odot$. A neutron-star nature of P is possible, but not
very likely. In any case, the adopted main-sequence nature of P is a rather
probable solution which is in
agreement with all observational constraints. Our
derived astrometric orbit does not depend sensitively on the nature of P,
since all the possible solutions indicate a very small value of $\beta$, so
that the difference between
the positions of A and of the photo-center of AP (see
Sect. 3.1.4) is small in any case.

Using ${\cal M}_{\rm A}$ and ${\cal M}_{\rm P}$
as derived above, the predicted semi-major
axis of the orbit of P relative to A is 142\,$\pm$\,21 mas. Our Table 7
provides a prediction of the position of P relative to A, if the ephemerides
for $\Delta x_{\rm{orb}, ph (AP)} (t)$
are multiplied by about $-$4.95. The
separation between A and P should be presently about 160 mas, is slightly
increasing to 186 mas until 2006, and is then decreasing to about 38 mas in
2017. Hence the next decade is especially favourable for resolving  the pair
$\alpha$ UMi AP. Of course, the large magnitude difference of more than 6$^m$
makes a direct observation of $\alpha$ UMi P rather difficult. Since A and P
seem to have nearly the same colour (as judged from the spectral types given in
Table 5), the magnitude difference should be (unfortunately) rather the same in
all the photometric bands.
Nevertheless we hope that modern interferometric techniques
or the use of other devices may be able to resolve the pair $\alpha$ UMi AP
during the next decade. Our paper provides hopefully
a fresh impetus for such investigations.

\section{Proper motion and position of Polaris}

\subsection{Center-of-mass of $\alpha$ UMi AP}


\begin{table*}

\caption[]{Proper motion $\mu_{\rm{cms} (AP)}$
and position $x_{\rm{cms} (AP)}$
of the center-of-mass of $\alpha$ UMi AP}

\tabcolsep1.3mm
\begin{tabular}{ccrrrrrrrrrrrrr}\hline\\[-1.5ex]
&&&&& \multicolumn{4}{c}{Prograde orbit} &&& \multicolumn{4}{c}{Retrograde
orbit}\\
&&&&&&&&&&&\multicolumn{4}{c}{(Preferred solution)}\\[0.5ex]\hline\\[-1.5ex]
Quantity & Unit & Epoch &&& in $\alpha_\ast$ & m.e. & in $\delta$ & m.e. &
&& in $\alpha_\ast$ & m.e. & in $\delta$ & m.e.\\[0.5ex]\hline\\[-1.5ex]
$\mu_{\rm{cms} (AP)}$
& [mas/year] & 1991.31 &&& + 41.17 & {\bf 0.50} & --
15.49 & {\bf 0.39} &&& + 40.81 & {\bf 0.50} & -- 15.22 & {\bf 0.39}\\
$\mu_{\rm{cms} (AP)}$
& [mas/year] & {\bf 1991.25} &&& {\bf + 41.17} & &
{\bf -- 15.49} &&&& {\bf + 40.81} & & {\bf -- 15.22} &\\
$\mu_{\rm{cms} (AP)}$ & [mas/year] & {\bf 2000.00} &&& {\bf + 41.17} &
& {\bf -- 15.50} &&&& {\bf + 40.81} & & {\bf -- 15.23} &\\[0.5ex]
\hline\\[-1.5ex]
$x_{\rm {ph(AP),av,H}}$ (+)
& [mas] & 1991.25 &&& 0.00 & 0.39 & 0.00 & 0.45 &&& 0.00
& 0.39 & 0.00 & 0.45\\
$x_{\rm {ph(AP),av,H}}$ (+)
& [mas] & 1991.31 &&& 0.00 & 0.39 & 0.00 & 0.45 &&& 0.00
& 0.39 & 0.00 & 0.45\\
$\Delta x_{\rm{orb}, ph (AP), av}$
& [mas] & 1991.31 &&& -- 10.37 & 2.67 &
+ 14.89 & 3.16 &&& + 16.15 & 3.08 & -- 8.58 & 2.78\\
$\Delta x_{\rm{mean \, ph} (AP)}$
& [mas] & 1991.31 &&& + 15.62 & 2.62 &
+ 12.40 & 3.54 &&& + 10.62 & 3.58 & + 17.05 & 2.58\\
$x_{\rm{cms} (AP)}$ (+)
& [mas] & 1991.31 &&& + 10.37 & {\bf 2.70} &
-- 14.89 & {\bf 3.19} &&& -- 16.15 & {\bf 3.11} & + 8.58 & {\bf 2.81}\\
$x_{\rm{mean} \, ph (AP)}$ (+)
& [mas] & 1991.31 &&& + 25.99 & 2.59 &
-- 2.49 & 4.35 &&& -- 5.53 & 4.27 & + 25.63 & 2.72\\[0.5ex]\hline\\[-1.5ex]
& &&&& \multicolumn{2}{c}{$\alpha$} &
\multicolumn{2}{c}{$\delta$} & & &
\multicolumn{2}{c}{$\alpha$} &
\multicolumn{2}{c}{$\delta$}\\[0.5ex]\hline\\[-1.5ex]
$x_{\rm {ph (AP), av, H}}$ & & 1991.25 &&&
\multicolumn{2}{r}{02$^h$ 31$^m$ 47\,.$\!\!^s$075254} &
\multicolumn{2}{r}{+ 89$^\circ$ 15' 50\,.$\!\!$"89698} &&&
\multicolumn{2}{r}{02$^h$ 31$^m$ 47\,.$\!\!^s$075254} &
\multicolumn{2}{r}{+ 89$^\circ$ 15' 50\,.$\!\!$"89698}\\
$x_{\rm {ph (AP), av, H}}$ & & 1991.31 &&&
\multicolumn{2}{r}{47\,.$\!\!^s$089026} &
\multicolumn{2}{r}{50\,.$\!\!$"89628} &&&
\multicolumn{2}{r}{47\,.$\!\!^s$089026} &
\multicolumn{2}{r}{50\,.$\!\!$"89628}\\
$x_{\rm{cms (AP)}}$ & & 1991.31 &&&
\multicolumn{2}{r}{47\,.$\!\!^s$142856} &
\multicolumn{2}{r}{50\,.$\!\!$"88139} &&&
\multicolumn{2}{r}{47\,.$\!\!^s$005192} &
\multicolumn{2}{r}{50\,.$\!\!$"90486}\\
$x_{\rm{mean \,ph (AP)}}$ & & 1991.31 &&&
\multicolumn{2}{r}{47\,.$\!\!^s$223939} &
\multicolumn{2}{r}{50\,.$\!\!$"89379} &&&
\multicolumn{2}{r}{47\,.$\!\!^s$060320} &
\multicolumn{2}{r}{50\,.$\!\!$"92191}\\[0.5ex]\hline\\[-1.5ex]
$x_{\rm{cms} (AP)}$ & & {\bf 1991.25} &&&
\multicolumn{2}{r}{{\bf 02$^h$ 31$^m$ 47\,.$\!\!^s$130034}} &
\multicolumn{2}{r}{{\bf + 89$^\circ$ 15' 50\,.$\!\!$"88232}} &&&
\multicolumn{2}{r}{{\bf 02$^h$ 31$^m$ 46\,.$\!\!^s$992482}} &
\multicolumn{2}{r}{{\bf + 89$^\circ$ 15' 50\,.$\!\!$"90578}}\\
$x_{\rm{cms} (AP)}$ & & {\bf 2000.00} &&&
\multicolumn{2}{r}{{\bf 02$^h$ 31$^m$ 48\,.$\!\!^s$999906}} &
\multicolumn{2}{r}{{\bf + 89$^\circ$ 15' 50\,.$\!\!$"74676}} &&&
\multicolumn{2}{r}{{\bf 02$^h$ 31$^m$ 48\,.$\!\!^s$846022}} &
\multicolumn{2}{r}{{\bf + 89$^\circ$ 15' 50\,.$\!\!$"77258}}\\[1.5ex]
$x_{\rm{cms} (AP)}$ & & {\bf 1991.25} &&&
\multicolumn{2}{r}{{\bf 37\,.$\!\!^\circ$94637514}} &
\multicolumn{2}{r}{{\bf + 89\,.$\!\!^\circ$26413398}} &&&
\multicolumn{2}{r}{{\bf 37\,.$\!\!^\circ$94580201}} &
\multicolumn{2}{r}{{\bf + 89\,.$\!\!^\circ$26414049}}\\
$x_{\rm{cms} (AP)}$ & & {\bf 2000.00} &&&
\multicolumn{2}{r}{{\bf 37\,.$\!\!^\circ$95416628}} &
\multicolumn{2}{r}{{\bf + 89\,.$\!\!^\circ$26409632}} &&&
\multicolumn{2}{r}{{\bf 37\,.$\!\!^\circ$95352509}} &
\multicolumn{2}{r}{{\bf + 89\,.$\!\!^\circ$26410349}}\\[0.5ex]\hline\\[-1ex]
\multicolumn{15}{l}{Explanation for (+): To the quantities marked with (+) in
the second part of Table 6, one has to add the HIPPARCOS positions
$x_{\rm{ph (AP), av, H}}$}\\
\multicolumn{15}{l}{at the corresponding epochs which are given in the first
two lines of the third part of Table 6.}
\end{tabular}

\end{table*}


The proper motion $\mu_{\rm{cms}(AP)}$ of the center-of-mass (cms) of the
closest components A and P of
$\alpha$ UMi has already been derived in Sect. 3.1.1 for
the epoch $T_{\rm{c, H}} = 1991.31$.
This proper motion is then transformed to the
other epochs by using strict formulae, assuming a linear motion of the cms of
AP in space and time. The values of $\mu_{\rm{cms}(AP)}$ for
the epochs 1991.25 and 2000.0 are given in Table 6.

In order to derive the position $x_{\rm{cms}(AP)}$ of the center-of-mass of
$\alpha$ UMi AP (Table 6), we first transform the HIPPARCOS position
$x_{\rm{ph (AP), av, H}}$ of the
photo-center of AP from epoch 1991.25 to $T_{\rm{c, H}} = 1991.31$ using
$\mu_{\rm{ph (AP), av, H}}$, since $T_{\rm{c, H}}$ corresponds
best to the effective mean epoch of the HIPPARCOS observations.
Then we subtract from $x_{\rm{ph (AP), av, H}} (1991.31)$ the
orbital displacements $\Delta x_{\rm{orb}, ph (AP), av} (1991.31)$, where
$\Delta x$ is calculated from the derived astrometric orbits (prograde and
retrograde), using the averaging method described by Eq. (17). This gives us
the position $x_{\rm{cms}(AP)} (1991.31)$ at the epoch $T_{\rm{c, H}}$.
Using the
proper motion $\mu_{\rm{cms} (AP)} (T_{\rm{c, H}})$,
we transform
$x_{\rm{cms} (AP)}$ from the epoch $T_{\rm{c, H}} = 1991.31$
to the standard epoch 2000.0.
For
the convenience of those users who like to use the HIPPARCOS standard epoch,
$T_{\rm H} = 1991.25$,
we give also the position $x_{\rm{cms}
(AP)}$ for this epoch $T_{\rm H}$. The values which should be used for
predicting the position
$x_{\rm{cms} (AP)} (t)$ and its mean error are given in
Table 6 in bold face. The right ascension $\alpha$ is given alternatively in the
classical notation (h, m, s) and, as done in the HIPPARCOS Catalogue, in
degrees and decimals of degrees. As discussed in Sect. 3.2.3, we propose to
use preferentially the retrograde orbit.

The position $x_{\rm{cms} (AP)} (t)$ at an arbitrary epoch
$t$ can be derived by using the strict formulae for epoch transformation, using
the epochs 2000.0 or 1991.25 as a starting epoch. The mean error
$\varepsilon_{x, {\rm{ cms}, (AP)}} (t)$
of $x_{\rm{cms} (AP)} (t)$ should be
derived from
\begin{eqnarray}
\lefteqn{
\varepsilon^2_{x, {\rm{cms (AP)}}} (t)
= \varepsilon^2_{x, {\rm{cms (AP)}}}
                                                (1991.31)}\nonumber\\[1ex]
& \hspace*{0.5cm} +\,\varepsilon^2_{\mu, {\rm{cms (AP)}}}
(t - 1991.31)^2 \,\,.
\end{eqnarray}                                                           
This equation assumes that $\mu_{\rm{cms} (AP)}$
and $x_{\rm{cms} (AP)}
(1991.31)$ are not correlated. This assumption is not strictly true. However,
for most applications it is not neccessary to allow for correlations, because
for epoch differences $\Delta t = |t - 1991.31|$ larger than a few years, the
second term in Eq. (20) is fully dominating. The correlation between
$\mu_{\alpha\ast, {\rm{cms (AP)}}}$
and $\mu_{\delta, {\rm{cms (AP)}}}$ is
negligably small (only caused by the tiny correlation between $\mu_{0,
\alpha\ast}$ and $\mu_{0, \delta}$).

All the quantities given in Table 6 refer to the HIPPARCOS/ICRS system
and to the equinox J2000 (but to various epochs).


\begin{table}

\caption[]{Orbital corrections
$\Delta x_{\rm{orb}, ph (AP)} (t)$ for the
photo-center of $\alpha$ UMi AP}

\tabcolsep1.85mm
\begin{tabular}{lrrrrr}\hline\\[-1.5ex]
& \multicolumn{2}{c}{Prograde orbit} & & \multicolumn{2}{c}{Retrograde orbit}\\
& & & & \multicolumn{2}{c}{(Preferred solution)}\\[0.5ex]\hline\\[-1.5ex]
Quantity  & in $\alpha_\ast$\hphantom{1} & in $\delta$\hphantom{1}
& & in $\alpha_\ast$\hphantom{1} & in $\delta$\hphantom{1}\\
or epoch $t$\\[0.5ex]\hline\\[-1.5ex]
$\Delta x_{\rm{orb}, \, ph (AP)} (t)$ & & & & &\\[0.5ex]
1987.00 & +   0.44 & --  7.89 & & --  7.97 & --  0.50\\
1988.00 & --  9.71 & --  3.30 & & --  2.18 & -- 10.08\\
1989.00 & -- 14.16 & +   3.53 & & +   5.19 & -- 13.71\\
1990.00 & -- 13.97 & +   9.49 & & +  11.15 & -- 12.81\\
1991.00 & -- 11.63 & +  14.10 & & +  15.51 & --  9.93\\
1992.00 & --  8.33 & +  17.62 & & +  18.65 & --  6.23\\
1993.00 & --  4.61 & +  20.27 & & +  20.88 & --  2.20\\
1994.00 & --  0.73 & +  22.23 & & +  22.39 & +   1.91\\
1995.00 & +   3.18 & +  23.61 & & +  23.33 & +   5.97\\
1996.00 & +   7.02 & +  24.51 & & +  23.78 & +   9.91\\
1997.00 & +  10.75 & +  24.99 & & +  23.83 & +  13.69\\
1998.00 & +  14.31 & +  25.09 & & +  23.51 & +  17.26\\
1999.00 & +  17.69 & +  24.86 & & +  22.89 & +  20.60\\
2000.00 & +  20.85 & +  24.32 & & +  21.98 & +  23.69\\
2001.00 & +  23.77 & +  23.51 & & +  20.82 & +  26.50\\
2002.00 & +  26.43 & +  22.43 & & +  19.43 & +  29.02\\
2003.00 & +  28.79 & +  21.11 & & +  17.83 & +  31.23\\
2004.00 & +  30.85 & +  19.56 & & +  16.03 & +  33.10\\
2005.00 & +  32.56 & +  17.79 & & +  14.06 & +  34.60\\
2006.00 & +  33.90 & +  15.81 & & +  11.92 & +  35.69\\
2007.00 & +  34.81 & +  13.64 & & +   9.64 & +  36.35\\
2008.00 & +  35.26 & +  11.29 & & +   7.22 & +  36.52\\
2009.00 & +  35.17 & +   8.77 & & +   4.70 & +  36.13\\
2010.00 & +  34.47 & +   6.09 & & +   2.10 & +  35.12\\
2011.00 & +  33.05 & +   3.29 & & --  0.55 & +  33.36\\
2012.00 & +  30.75 & +   0.41 & & --  3.17 & +  30.73\\
2013.00 & +  27.37 & --  2.48 & & --  5.68 & +  27.02\\
2014.00 & +  22.60 & --  5.24 & & --  7.89 & +  21.92\\
2015.00 & +  15.96 & --  7.53 & & --  9.42 & +  15.03\\
2016.00 & +   6.90 & --  8.57 & & --  9.40 & +   5.86\\
2017.00 & --  4.13 & --  6.54 & & --  6.08 & --  4.90\\[1.5ex]
1987.66 \, Periastron & --  6.71 & --  5.33 & & --  4.57 & --  7.33\\
2002.46 \, Apastron &  +  27.54 & +  21.86 & & +  18.73 & +  30.07\\[1.5ex]
1991.31 \, instantan. &  -- 10.67 & +  15.30 & & +  16.60 & --  8.84\\
1991.31 \, averaged &  -- 10.37 & +  14.89 & & +  16.15 & --  8.58\\
[0.5ex]\hline\\[-1.8ex]
$\Delta x_{\rm{mean} \, ph (AP)}$
& + 15.62 & + 12.40 & & + 10.62 & + 17.05\\[0.5ex]
(1991.31) & & & & &  \\[0.5ex]\hline
\end{tabular}

\end{table}


\subsection{Orbital corrections for the photo-center of $\alpha$ UMi AP}

In order to obtain a prediction for the instantaneous position
$x_{\rm{ ph (AP)}} (t)$
of the photo-center of $\alpha$ UMi AP at an epoch $t$, one has to add the
orbital correction
$\Delta x_{\rm{orb, ph (AP)}} (t)$ to the position of the
center-of-mass $x_{\rm{cms (AP)}} (t)$:
\begin{equation}
x_{{\rm ph (AP)}} (t)
= x_{\rm{cms (AP)}} (t) + \Delta x_{\rm{orb, ph (AP)}} (t)
                                                          \,\,.        
\end{equation}
The ephemerides for the orbit of the photo-center of AP are given in Table 7.
The orbital elements used in calculating the ephemerides are those listed in
Table 4. Usually it is allowed to neglect the effect that the $\alpha\delta$
system is slightly rotating ($\dot{\Theta} = + 0\,.\!\!^\circ00088$/year), due
to the motion of Polaris
on a great circle. Table 7 lists also the position of the
intantaneous photo-center at periastron, apastron,
and at $T_{\rm{c, H}}$. The small
difference between the instantaneous position and the averaged position (Sect.
3.1.5) of the photo-center at $T_{\rm{c, H}}$ shows that the deviations of the
fitting straight line from the actual orbits remain mostly below 1 mas within
the interval of
$D_{\rm H} = 3.1$ years of the HIPPARCOS observations, since these
deviations reach their maximum at the borders of $D_{\rm H}$,
namely about twice the
deviation at $T_{\rm{c, H}}$. The very small deviations from a straight line
explain
also why we were, during the HIPPARCOS data reduction, unable to obtain an
orbital (O) solution or an acceleration (G) solution for Polaris, although we
tried to do so.

At the end of Table 6, we give the (constant) off-set
between the mean photo-center and the center-of-mass. All values are valid for
the equinox J2000.0, and for the orientation of the $\alpha\delta$ system at
epoch 1991.31 (which differs from that at epoch 2000.0 by $\Delta\Theta =
- 0\,.\!\!^\circ008$ only).

The typical mean error of $\Delta x_{\rm{orb}, ph (AP)}$,
due to the
uncertainties in the orbital elements (mainly in $\Omega$), is about $\pm$\,5
mas. It varies, of course, with the orbital phase,
approximately between $\pm$\,2 mas and $\pm$\,7 mas.
However, a detailed
calculation of this mean error is often unnecessary for deriving the mean
error of the prediction for $x_{{\rm ph (AP)}} (t)$,
since the mean error of
$x_{\rm{ph (AP)}} (t)$
is governed by the mean error of
$\mu_{\rm{cms} (AP)}$ for epoch
differences $|t - T_{\rm{c, H}}|$ larger than about 20 years.


\begin{table}

\caption[]{Comparison between predicted positions
for the photo-center of $\alpha$ UMi AP}

\tabcolsep1.81mm
\begin{tabular}{rrrrrrr}\hline\\[-1.5ex]
& & \multicolumn{2}{c}{Prograde orbit} & & \multicolumn{2}{c}{Retrograde
orbit}\\
& & & & & \multicolumn{2}{c}{(Preferred solution)}\\[0.5ex]
Difference & Epoch & in $\alpha_\ast$  & in $\delta$ & & in $\alpha_\ast$ &
in $\delta$\\
& $t$\hphantom{00} & \multicolumn{2}{c}{[mas]} & &
\multicolumn{2}{c}{[mas]}\\[0.5ex]\hline\\[-1.5ex]
\multicolumn{7}{l}{HIPPARCOS prediction minus this paper (instantaneous
position):}\\[0.5ex]
& 1900.00 & -- 275 & -- 330 & & -- 298 & -- 313\\
$T_{\rm{c, H}}$ \hspace{0.30 cm} = & 1991.31 & --   0 &  +   0 & &  +   0 & --
0\\
& 2000.00 & --   5 &  +  23 & &  +  24 & --   2\\
& 2010.00 &  +  12 &  +  79 & &  +  78 &  +  21\\
& 2020.00 &  +  90 &  + 111 & &  + 101 &  + 103\\[0.5ex]\hline\\[-1.5ex]
\multicolumn{7}{l}{FK5 minus this paper (mean photo-center):}\\[0.5ex]
& 1900.00 & --   1 & --  14 & & --   2 & --  18\\
$T_{\rm{c}, \delta, \rm{FK5}}$ = & 1916.08 &  +  4 & --  34 & &  +  9 & -- 42\\
$T_{\rm{c}, \alpha, \rm{FK5}}$ = & 1927.17 &  +  8 & --  47 & &  + 16 & -- 58\\
& 1991.31 &  + 29 & -- 126 & &  + 60 & -- 154\\
& 2000.00 &  + 31 & -- 137 & &  + 66 & -- 167\\
& 2010.00 &  + 35 & -- 149 & &  + 73 & -- 182\\
& 2020.00 &  + 38 & -- 162 & &  + 80 & -- 197\\[0.5ex]\hline\\[-1.5ex]
\multicolumn{7}{l}{Prograde minus retrograde orbit (instantaneous
positions):}\\[0.5ex]
& 1900.00 & -- 23 & +  17\\
& 1991.31 & --  1 & +   1\\
& 2000.00 & +  29 & -- 25\\
& 2010.00 & +  66 & -- 56\\
& 2020.00 & +  11 & --  8\\[0.5ex]\hline
\end{tabular}

\end{table}


\subsection{Comparison of positions}

In Table 8 we compare positions predicted by our results with those predicted by
HIPPARCOS and by the FK5.

At epoch $T_{\rm{c, H}} = 1991.31$, the positions of the photo-center of AP
predicted by our results (for both types of orbits) agree with the HIPPARCOS
position by construction (except for the slight difference between the
instantaneous and averaged position).

From Fig. 1 we see that the HIPPARCOS predictions for small epoch differences
$\Delta t = |t - T_{\rm{c, H}}|$,
say for $\Delta t < 4$ years, are also in good
agreement with our predictions, since the HIPPARCOS data are essentially a
tangent to our astrometric orbits. In other words, the HIPPARCOS data are a
good short-term prediction
(relative to $T_{\rm{c, H}}$) in the terminology of
Wielen (1997). For larger epoch differences (Table 8), the HIPPARCOS prediction
for $x_{{\rm ph (AP)}} (t)$
starts to deviate significantly from our predictions.
Going to the past, e.g. to $t = 1900$, the differences reach large values of
about 300 mas = 0\,.$\!\!$"3 in each coordinate. Such differences are already
larger than the measuring errors of some meridian circles at that time,
especially for
Polaris.
(The formal mean errors in $\alpha_\ast$ and $\delta$
of the position predicted by the linear HIPPARCOS solution
at the epoch 1900 are 43 mas and 50 mas only.)
The reason for the failure of a linear prediction based directly on
the HIPPARCOS Catalogue is the fact that the quasi-instantaneously measured
HIPPARCOS proper motion of Polaris contains an orbital motion
$\Delta\mu_{\rm{tot}}$ of about 5 mas/year as a `cosmic error'.

Our data reproduce rather well the FK5 positions at the central FK5 epochs.
This is to be expected, since we have made use of these positions in
determining $\mu_0$ (and hence $\mu_{\rm{cms}}$).
For $T_{\rm{c, H}} = 1991.31$, the
FK5 prediction deviates rather strongly from our values. This is
in accordance with the mean error of $\mu_{\rm{FK5}}$ and the large epoch
differences $T_{\rm{c, H}} - T_{\rm{c, FK5}}$.
(The mean errors
in $\alpha_\ast$ and $\delta$
of the FK5 position (reduced to the HIPPARCOS system) are
at the central epochs (given in Table 8) 37 mas and 34 mas,
and at epoch 1991.31 72 mas and 66 mas.)

How large are the differences between the positions which we predict if we
use
either the retrograde orbit or the prograde one\,?
At $T_{\rm{c, H}} = 1991.31$, the
differences are nearly zero by construction. At other epochs, the
orbital differences can be seen in Fig. 1. To these differences in
the orbital corrections, we have to add the slight positional differences
which are due to the differences in $\mu_{\rm{cms}}$ of both orbits.
The total differences between the prograde and retrograde orbit are shown at
the end of Table 8 for some epochs. An extremum in these differences occurs in
$\alpha_\ast$ (+ 68 mas) and in $\delta$ (-- 59 mas) at about the year 2012.

\subsection{Space velocity of Polaris}

From the derived proper motions $\mu_{\rm{cms} (AP)}$ of the center-of-mass of
$\alpha$ UMi AP (Table 6, retrograde orbit),
from the radial velocity $v_{\rm r} = \gamma$ (Table 4), and
from the
HIPPARCOS parallax
$p_{\rm H}$ (Eq. 14), we derive the three components $U$, $V$, $W$
of the space velocity {\bf v} of Polaris (Table 9). We neglect a possible
intrinsic K term in the pulsating atmosphere of the Cepheid $\alpha$ UMi A
(Wielen 1974). This is probably justified, especially in view of the very small
amplitude of the radial velocity due to pulsation.


\begin{table}

\caption[]{Space velocity of the center-of-mass of $\alpha$ UMi AP. For
detailed explanations see Sect. 4.4\,.}

\tabcolsep1.11mm
\begin{tabular}{lrrrrrrrr}\hline\\[-1.5ex]
& \multicolumn{4}{c}{Prograde orbit}  & \multicolumn{4}{c}{Retrograde
orbit}\\
& & & & & \multicolumn{4}{c}{(Preferred
solution)}\\[0.5ex]\hline\\[-1.5ex]
Velocity & $U$ & $V$ & $W$ & $v$  & $U$ & $V$ & $W$ & $v$\\
& \multicolumn{4}{c}{[km/s]}  &
\multicolumn{4}{c}{[km/s]}\\[0.5ex]\hline\\[-1.5ex]
{\bf v}$_{\rm{S0}}$ & -- 14.4 & -- 28.2 & -- 5.5 & 32.1  & -- 14.2 & -- 28.0
& -- 5.4 & 31.9\\
m.e. & $\pm$ 1.2 & $\pm$ 0.8 & $\pm$ 1.0 & &  $\pm$ 1.2 & $\pm$ 0.8 & $\pm$
1.0\\[1.5ex]
{\bf v}$_{\rm{L0}}$ & -- 5.4 & -- 16.2 & + 1.5 & 17.1  & -- 5.4 & --
16.0 & + 1.6 & 17.0\\[1.5ex]
{\bf v}$_{\rm{C0}}$ & -- 8.0 & -- 16.0 & + 1.5 & 18.0  & -- 8.0 & -- 15.8 &
+ 1.6 & 17.8\\[0.5ex]\hline
\end{tabular}

\end{table}


The velocity component $U$ points {\it towards} the galactic center, $V$ in the
direction of galactic rotation, and $W$ towards the galactic north pole. The
velocity {\bf v}$_{\rm{S0}}$ is measured relative to the Sun. The velocity
{\bf v}$_{\rm{L0}}$ refers to the local standard of rest. For the solar
motion we use {\bf v}$_\odot$ = (+\,9, +\,12, +\,7) km/s, proposed by
Delhaye (1965). The velocity {\bf v}$_{\rm{C0}}$ is the peculiar velocity
of Polaris with respect
to the circular velocity at the position of Polaris (see
Wielen 1974). For the required Oort constants of galactic rotation, we adopt
$A
= +14$ (km/s)/kpc and $B = -12$ (km/s)/kpc. As mentioned in Sect. 2.3, the
velocity of the center-of-mass of $\alpha$ UMi AP may differ from that of
$\alpha$ UMi AP+B by a few tenth of a km/s.

The peculiar velocity {\bf v}$_{\rm{C0}}$ of Polaris is reasonable for a
classical Cepheid. According to Wielen (1974), the velocity dispersions
($\sigma_U$, $\sigma_V$, $\sigma_W$) for nearby classical
Cepheids are (8, 7, 5) km/s. Hence only
the $V$ component of {\bf v}$_{\rm{C0}}$ of Polaris is slightly larger than
expected on average.



\end{document}